\documentclass{article}
\usepackage{booktabs} 
\usepackage{float}
\usepackage{graphicx}
\usepackage{url}

\begin{document}
\begin{center}
\fbox{\parbox{0.96\linewidth}{
\small
© 2024 Association for Computing Machinery. 
This is the author's version of the work. It is posted here by permission of ACM for your personal use. 
Not for redistribution. The definitive version of record will be published in:
\textit{Proceedings of the ACM International Conference on Research in Adaptive and Convergent Systems (RACS '24), November 5–8, 2024, Pompei, Italy}.\\
DOI: [To be inserted upon publication]
}}
\end{center}


\title{Extracting Information from Scientific Literature via Visual Table Question Answering Models}

\author{
Dongyoun Kim$^{1}$\thanks{dongyoun.kim@jacks.sdstate.edu} \and
Hyung-do Choi$^{2}$\thanks{choihd@etri.re.kr} \and
Youngsun Jang$^{3}$\thanks{youngsunjang@muhlenberg.edu} \and
John Kim$^{4}$\thanks{jkim@utica.edu}
}

\date{
$^{1}$South Dakota State University, Brookings, SD, USA \\
$^{2}$Electronics and Telecom. Research Institute, Daejeon, South Korea \\
$^{3}$Muhlenberg College, Allentown, PA, USA \\
$^{4}$Utica University, Utica, NY, USA \\
}

\maketitle

\begin{abstract}
This study explores three approaches to processing table data in scientific papers to enhance extractive question answering and develop a software tool for the systematic review process. The methods evaluated include: (1) Optical Character Recognition (OCR) for extracting information from documents, (2) Pre-trained models for document visual question answering, and (3) Table detection and structure recognition to extract and merge key information from tables with textual content to answer extractive questions. In exploratory experiments, we augmented ten sample test documents containing tables and relevant content against RF- EMF-related scientific papers with seven predefined extractive question-answer pairs. The results indicate that approaches preserving table structure outperform the others, particularly in representing and organizing table content. Accurately recognizing specific notations and symbols within the documents emerged as a critical factor for improved results. Our study concludes that preserving the structural integrity of tables is essential for enhancing the accuracy and reliability of extractive question answering in scientific documents.

\end{abstract}


%
%




\maketitle

\section{Introduction}
The systematic review of scientific literature is a cornerstone of evidence-based research, enabling researchers to synthesize existing knowledge and draw meaningful conclusions from various studies. As the volume of scientific publications grows, a need for efficient and accurate tools to assist in the review process becomes increasingly critical \cite{won2022design}. In particular, our systematic review framework aims to evaluate related literature by using a question-answering approach on the associations between exposure to radiofrequency electromagnetic fields (RF-EMF) and oxidative stress \cite{RF}. The global population is increasingly exposed to radiofrequency electromagnetic fields (RF-EMF), with future exposure levels expected to rise further. Growing concerns about the potential public health implications of RF-EMF highlight the urgent need for comprehensive health risk assessments to inform the public. However, determining the specific levels of EMF that significantly affect human health remains challenging, as no universally accepted standards exist for assessing EMF-related health risks \cite{Verbeek2021, Jeong2021, Schuermann2021, RF}.

\begin{figure}[H]
\includegraphics[width=0.9\textwidth]{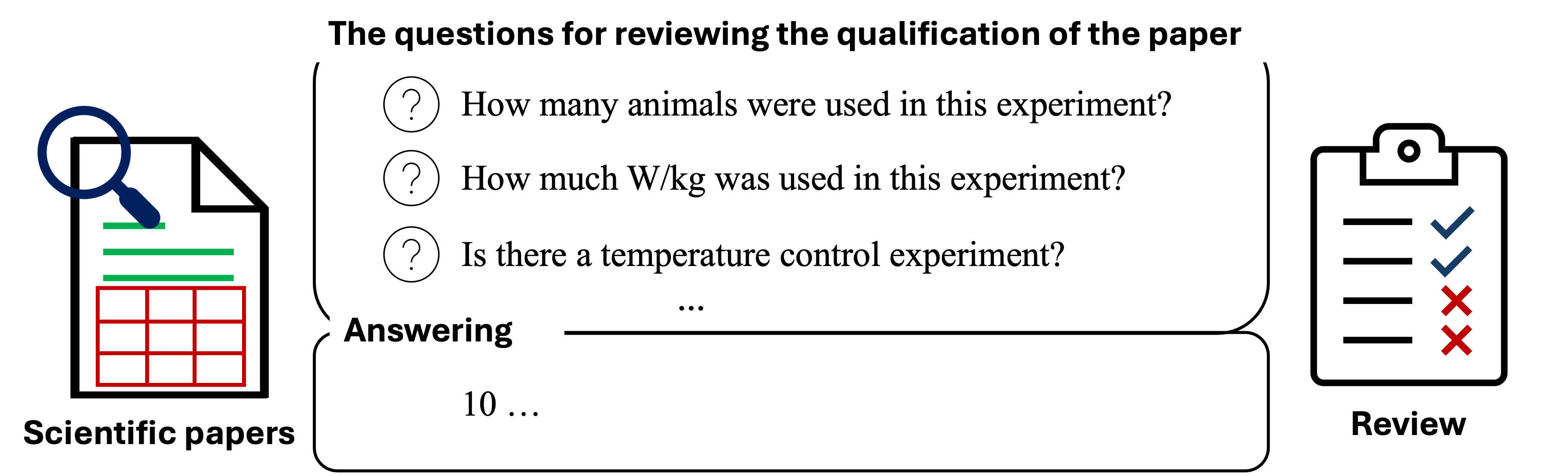}
\caption{Overview of the Systematic Review of Scientific Literature}
\label{Fig0}
\end{figure}

Computer Vision (CV) and Natural Language Processing (NLP) techniques have been instrumental in automating the classification of scientific papers and extracting relevant information, thereby streamlining the systematic review process. Figure ~\ref{Fig0} shows an overview of the reviewing system. While the latest large language models fine-tuned on scientific literature have achieved high accuracy in extracting relevant text-only paragraph information, challenges persist in extracting information from figures and tables in it \cite{won2022design,mathew2021docvqa}. 

To address these challenges, A significant amount of research has been carried out in document understanding aimed at preserving the layout information of documents \cite{pack2023weakly}, particularly for improving the interpretation of complex tables and figures \cite{kim2022ocr, mathew2021docvqa, ding2022vdoc,lee2023pix2struct}. In the context of systematic reviews, table data is crucial, as it often represents complex structured data formats containing key information. Extracting and interpreting this information from visual document formats is challenging due to the need to understand the complex structured data in context—within the table’s rows and columns and in relation to the surrounding text \cite{smock2022pubtables}. Furthermore, scientific papers frequently involve a variety of notations and symbols, adding additional complexity \cite{won2022design}.

In this paper, we explore three approaches to document visual question-answering (DocQA) aimed at handling table data to enhance the performance of extractive question answering $—$ a crucial task in systematic reviews: (1) Optical Character Recognition (OCR)-based methods that convert table images into machine-readable text, such as plain text or LaTeX format; (2) end-to-end vision-language models that directly interpret document images for question-answering; and (3) table detection and structure recognition models that preserve the integrity of tabular data from visual documents while integrating it with surrounding text for comprehensive analysis. To evaluate the capabilities of systematic reviewing, we modified ten sample scientific documents on radio-frequency electromagnetic fields (RF-EMF) to include tables and related contextual information. We then compared the capabilities of these methods in extractive question-answering tasks using a set of seven predefined question-answer pairs focused on experimental subjects, exposure levels, and statistical validity. These predefined question-answer pairs were defined to make it easier for the large language model to understand by generating simple questions based on prior research, particularly the work of B. Henschenmacher et al. \cite{RF}.

Overall, our contributions in this work are summarized as follows:
\begin{itemize}
\item We exploit three distinct methodologies for scientific visual papers, focusing on enhancing extractive question-answering performance.
\item We demonstrate how each approach handles the unique challenges of extracting and interpreting structured data from tables, identifying each method's strengths and limitations in preserving structural information from visual documents.
\end{itemize}
The remainder of this paper is organized as follows. Section $2$ reviews related work in document visual question answering, table detection, and OCR-based methods. Section $3$ describes our methodology in detail, outlining the approaches and experimental setup. Section $4$ presents the experimental results and discusses the findings. Finally, Section $5$ concludes with recommendations for future work, emphasizing the need for fine-tuning models and enhancing OCR integration to improve the accuracy and reliability of extractive question-answering tasks in systematic reviews.

\section{Related work}
\subsection{Document Visual Question Answering}
Document Visual Question Answering (DocQA) is an emerging field at the intersection of information extraction tasks and reasoning for generating answers from question queries. It's an extension of the Visual Question Answering (VQA) tasks \cite{antol2015vqa}, so the question-answer pairs of DocQA involve extracting information not only from interpreting images but also from the text, layout, and structural elements within the document image.
Early DocQA systems primarily relied on Optical Character Recognition (OCR) technologies to extract text from images. These methods converted visual information into machine-readable text, which was then processed using natural language processing (NLP) techniques to answer queries. Systems such as Tesseract \cite{smith2007overview} were foundational in OCR technology. However, OCR-based tasks highly rely on their performance \cite{kim2022ocr, lee2023pix2struct}. As a result, the effectiveness of most DocQA models is often constrained by the accuracy and reliability of the OCR component, especially when dealing with degraded images, complex layouts, or documents containing non-textual elements like tables and mathematical symbols.

There has been a rise of multimodal models capable of processing texts as well as images. Kim et al. \cite{kim2022ocr} proposed the OCR-free Document Understanding Transformer (Donut) model that processes document images end-to-end without explicit OCR, directly generating sequences of text that answer specific questions. This model has shown particular strengths in handling tabular data and structured information, although it sometimes struggles with the interpretation of complex symbols and layouts. Pix2Struct \cite{lee2023pix2struct}, on the other hand, focuses on interpreting document images by learning their structure directly from pixels. This approach enables it to understand and preserve the spatial relationships in documents, which is particularly useful for maintaining the integrity of tables and forms. However, as in other models, Pix2Struct faces challenges in accurate parsing and understanding a wide variety of symbols and complex structures, particularly when dealing with noisy or low-resolution images. GPT-4 \cite{chatgpt2024}, a multimodal large language model, has demonstrated a definite promise in multimodal understanding and reasoning in various fields by leveraging extensive pre-training on large-scale multi-modal datasets, including texts and images. In question-answering tasks, GPT-4's ability to generate responses from complex documents is enhanced by its capability to interpret not only textual content but also the contextual and structural nuances of documents \cite{achiam2023gpt4}.

Despite these advancements, significant challenges persist in DocQA, particularly in handling a diverse range of document layouts and types. Another crucial issue is contextual and structural information preservation, especially when dealing with complex tables and forms.

This paper presents examples of challenge cases and evaluates the effectiveness of pre-trained DocQA models using predefined question-answer pairs for classifying quality scientific documents on radio-frequency electromagnetic fields involving complex symbols and structures.

\subsection{Table Detection and Table Structural Recognition}
An application of OCR allows researchers to recognize and digitize text from documents accurately. When combined with table structure recognition, OCR enhances the overall functionality and usability of document processing. Nevertheless, recognizing tables from unstructured documents presents challenges due to varied structures and styles of tabular data. The complexity and diversity in table formats require advanced techniques and algorithms to achieve accurate and efficient table extraction, which is a critical component in comprehensive document analysis and processing.
Smock et al. \cite{smock2022pubtables} proposed the PubTables-1M dataset, containing 948K table images with corresponding annotations in an attempt to advance table recognition in document images. Their work focuses on improving the extraction of complex table structures by providing a large-scale dataset that supports the development of more robust models. Raja et al. \cite{raja2022visual} focused on recognizing complex table structures from document images, proposing a novel object-detection-based deep model and a rectilinear graph-based formulation to improve structure recognition. They emphasized the importance of empty cells in tables and introduced an enhanced evaluation criterion.

This paper presents examples of challenge cases in the DocQA task, focusing on the difficulties associated with table detection and structural recognition. By analyzing these cases, we aim to show the limitations of current models and propose potential solutions to enhance the accuracy and reliability of table structure recognition in document-based question answering.

\subsection{Optical Character Recognition (OCR)}
The increasing focus on information retrieval has driven extensive research in OCR to extract text from both structured and unstructured documents, thereby playing an essential role in document processing. It involves converting scanned images of text into machine-encoded text, allowing for easier editing, searching, and storage of documents. Various studies have highlighted the importance of OCR in document analysis and recognition \cite{hamad2016detailed} Despite its benefits, OCR technology still faces challenges, such as complex layouts, degraded images, or a loss of structural information, which limits its effectiveness \cite{kim2022ocr, gurgurov2024image}.

This paper demonstrates examples of these challenges in degraded images. It presents comparative studies of text OCR and LaTeX OCR \cite{gurgurov2024image} to mitigate the loss of structural information in DocQA tasks. By analyzing the performance of these OCR models, we aim to provide insights into their strengths and limitations, particularly in the context of document-based question answering, where preservation of document structure is essential for accurate interpretation and response generation.

\section{methodology}
\begin{figure}
\includegraphics[width=0.9\textwidth]{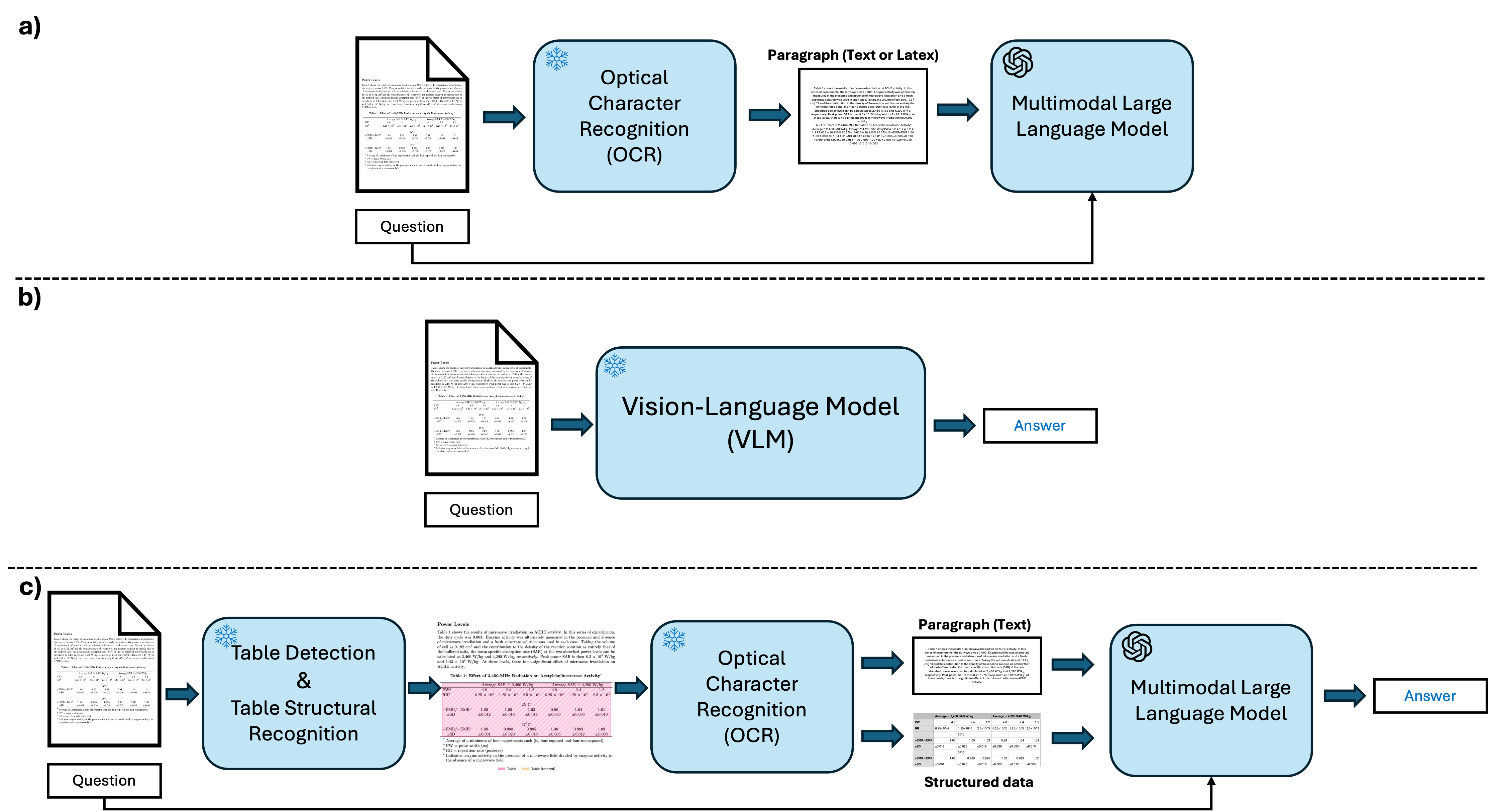}
\caption{Comparative Study Design for Document Visual Question-Answering Approaches, a) OCR-based, b) Vision- Language Models, and c) Table Structure-Aware Approach.}
\label{Fig1}
\end{figure}

As illustrated in Figure ~\ref{Fig1}, we explore three approaches for the scientific DocQA task: (1) OCR-based approach: extraction of text information from OCR + the Question-Answer module from GPT- 4 API (2) End-to-end vision-language model: pre-trained models for document question-answering task, and (3) Table-structure- aware approach: Transformer-based Table Detection and Structural Recognition + OCR +the Question-Answer module from GPT-4 API.

\subsection{OCR-based approaches}
As a prior study, the OCR-based approach for DocQA tasks is an intuitive way to understand visual documents by converting visual content into text format. The primary objectives of this study are twofold: 1) to visually validate the capabilities of text OCR and LaTeX OCR models, specifically Tesseract and Pix2tex \cite{lee2023pix2struct}, in specific scientific documents, and 2) to assess the impact of these DocQA tasks when using pseudo-text labels and pseudo-LaTeX labels generated by the respective OCR models.

In the first study, the text OCR might extract text information in the document, but it is predicted to face common challenges such as temperature degree symbols, mathematical notation, and a tabular data structure in complex scientific papers. On the other hand, LaTeX OCR is expected to not only represent complex symbols but also maintain tabular data structure.

In the second study, assuming these OCR models would perform well, the QA module generates a response from the pseudo-label in order to evaluate the capabilities of the document representation formats, such as text with escape characters and LaTex grammar. A detailed discussion of these approaches is discussed in section $5.1$

\subsection{End-to-End Visual-Language Model}
In contrast to traditional OCR-based approaches, this study analyzes the use of end-to-end large-scale pre-trained models for Document Question Answering (DocQA) tasks. Specifically, we explore the capabilities of three prominent models: Donut \cite{kim2022ocr}, Pix2Struct \cite{lee2023pix2struct}, and GPT-4 \cite{chatgpt2024}. 

Donut is an OCR-free, transformer-based model designed to process document images directly, bypassing the need for text extraction via OCR. This model leverages the visual information inherent in document images, which allows it to maintain and utilize the layout and structural cues crucial for understanding complex documents.
Pix2Struct is a model pre-trained to convert visual data(e.g., screenshots) into a semi-structured format such as HTML. This conversion process facilitates the interpretation and analysis of the document's structure, enabling the model to retain and utilize the structural integrity of the document.
As one of the leading models for question reasoning and multimodal processing, GPT-4 integrates both visual and textual inputs to generate sophisticated responses to DocQA tasks. By directly processing images while maintaining their layout, GPT-4 is capable of understanding and reasoning about the complex, symbolic, and structural elements often found in scientific documents. This model's ability to combine text understanding with visual layout information makes it a powerful tool for comprehensive document analysis.

We utilize pre-trained weights for each model and conduct a series of experiments across specific scientific document question-answering tasks. The tasks are designed to assess the models' ability to handle documents with different levels of complexity, including those with intricate layouts, complex symbolic representations, and diverse structural elements. Detailed results of these approaches are discussed in section $5.2$.

\subsection{Table structure-aware approach}
The core concept behind the table structure-aware approach is to enhance question-answering models by enabling them to better represent visual documents containing tabular data. This is achieved by using table structure recognition models specifically designed to identify and preserve the complex structures inherent in tables. These models can more accurately interpret table structure by maintaining the integrity of rows, columns, and cells.

We follow the inference pipeline of table transformers \cite{smock2022pubtables}. This finds column/row structures using transformer-based object detection and extracts the text from each cell region using OCR. Due to the multiple stages,  the question-answering module is subject to accuracy at various stages of the processing pipeline, which might also be reduced.  A detailed discussion of this approach is discussed in section 5.3 

\section{experiment}
\subsection{Dataset}
We evaluate these approaches on ten scientific documents with a table in the field of Radio Frequency Electromagnetic Field (RF-EMF) that were previously analyzed  \cite{won2022design, liebl2021evaluation, dilin2016scribblesup, xu2017page, RF}. The ten scientific documents involve the context consisting of the table in the part of the dataset \cite{won2022design,RF}, as shown in Figure 2. Additionally, we augment the enhanced-quality image, LaTeX, and text format document data to avoid performance degradation primarily caused by document quality issues. The question-answer pairs are derived from human-written lists for paper-quality classification in the field of RF-EMF \cite{won2022design}, as detailed in Table ~\ref{tab1}.

\begin{figure}
\includegraphics[width=0.9\textwidth]{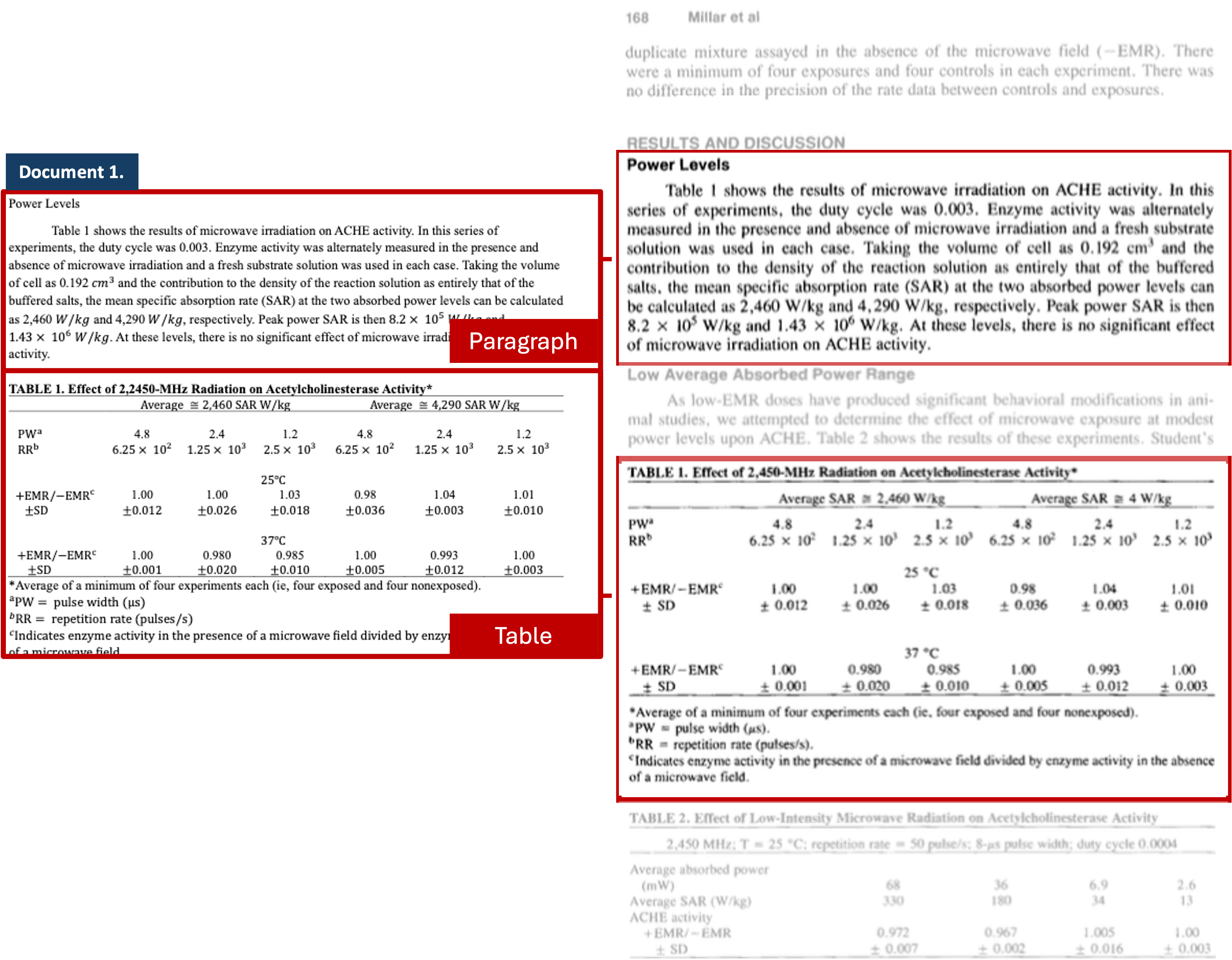}
\caption{Example of a Modified Document with Integrated Tables and Contextual Paragraphs for Extractive Question Answering Model.}
\label{Fig2}
\end{figure}

\begin{table}
  \caption{Pre-defined Question Templates for Extractive Question Answering in RF-EMF Experiments.}
  \label{tab1}
  \begin{tabular}{p{0.43\linewidth} p{0.43\linewidth}}
    \toprule
    Question Number/Area & Question\\
    \midrule
    $Q1)$ Experiment subject & What animal has been used?\\
    $Q2)$ Number of the subjects & How many animals were used?\\
    $Q3)$ Specific Absorption Rate (SAR) & How much $W/kg$ was used?\\
    $Q4)$ RF frequency & What is the signal frequency?\\
    $Q5)$ Validity evaluation & Is the experiment statistically valid (significant)?\\
    $Q6)$ Temperature control & Is there a temperature control experiment?\\
    $Q7)$ Generous information of experiment & What are the variables being compared in this table?\\
    \bottomrule
  \end{tabular}
\end{table}

\subsection{Evaluation method}
We define specific evaluation cases to assess the performance of these approaches, as detailed in Table ~\ref{tab3}. These cases are necessary because the predefined questions demand different capabilities, such as understanding paragraphs/tables and interpreting mathematical symbols, if any, included in each document.

\begin{table}
  \caption{Evaluation Criteria for Answering Capabilities in Extractive Question Answering Tasks.}
  \label{tab3}
  \begin{tabular}{p{0.2\linewidth} p{0.3\linewidth} p{0.4\linewidth}}
    \toprule
    Evaluation Case & Answer Capability & Answer Correctness\\
    \midrule
    Case 1 & Retrieve the answer a. From paragraph b. From table c. No Answer (Q1-Q7) & Correct: If the keyword is correctly retrieved. Partial Correct: If only part of the answer is retrieved. Wrong: If a hallucinated or incorrect answer is returned.\\
    Case 2 & Understanding Mathematical Symbol Meaning(Q2, Q3, Q4, Q5, Q6) & Correct:
If the exact number is returned.
Partial Correct:
If an incorrect number or only part of the answer is returned.\\
    Case 3 & Understanding table structure (Q7)& Correct:
If columns or rows data are correctly returned.\\
  \bottomrule
\end{tabular}
\end{table}

\section{RESULTS AND DISCUSSION}
\subsection{OCR-based approach}
\begin{figure}
\includegraphics[width=0.9\textwidth]{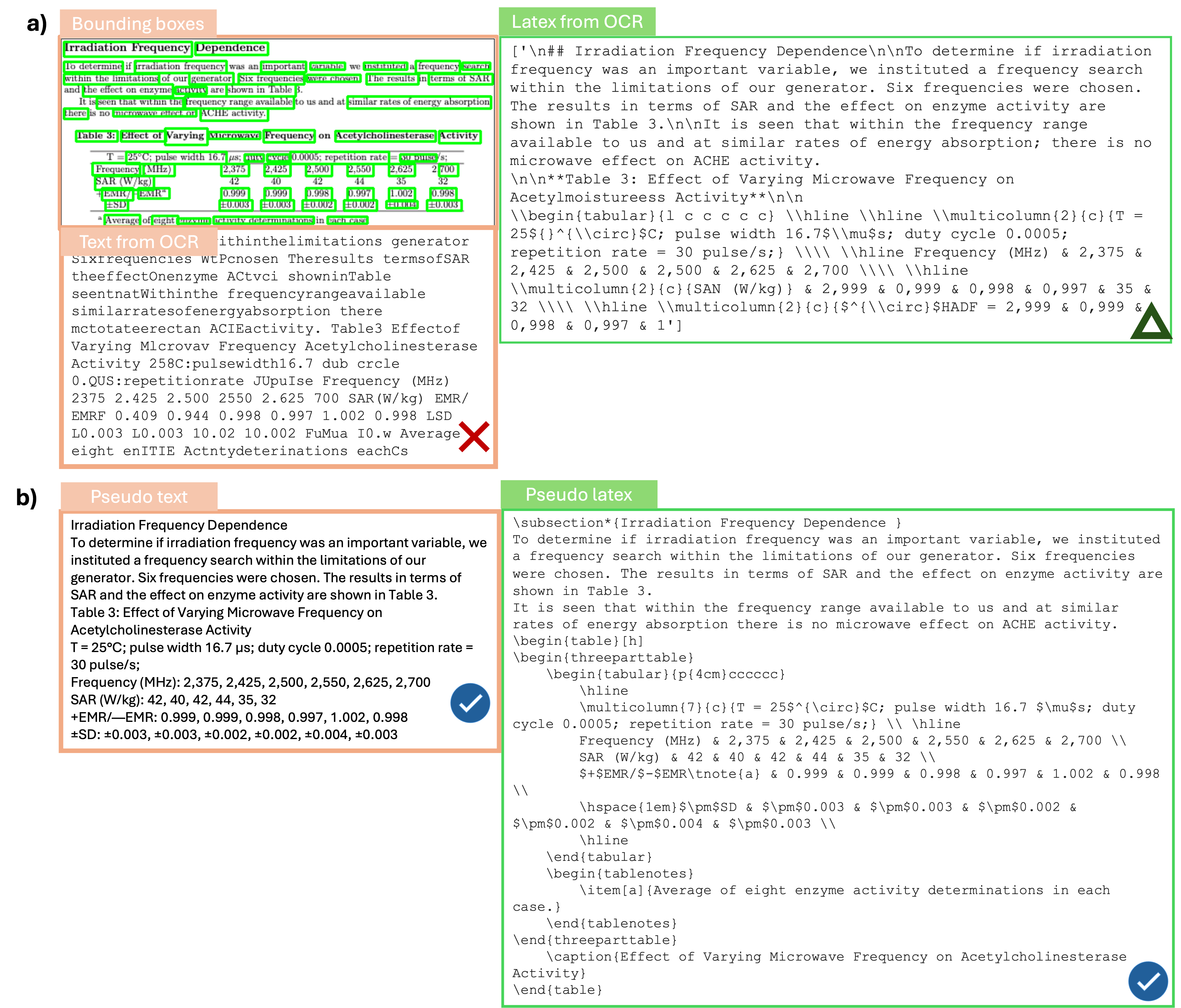}
\caption{a) OCR Model Results, b) Comparison with Pseudo- Text Outputs.}
\label{Fig3}
\end{figure}

To evaluate the capabilities of OCR-based approaches, we present preliminary experiments with sample results from OCR models, as shown in Figure ~\ref{Fig3}a. The results indicate that OCR-based approaches are affected by the complexity of documents containing specific symbols, as observed in our dataset. Therefore, we provide pseudo-linearized text data, as illustrated in Figure ~\ref{Fig3}b. Despite offering enhanced-quality images to the OCR model, these approaches still struggle to represent exponential symbols, temperature symbols, and structural information. To mitigate this limitation, we use a prompt that informs the answer generation module that the input data is in a linearized text, such as: “From this linearized text content (or Latex format content), question: {}. Provide the result from the question.”

\begin{figure}
\includegraphics[width=0.9\textwidth]{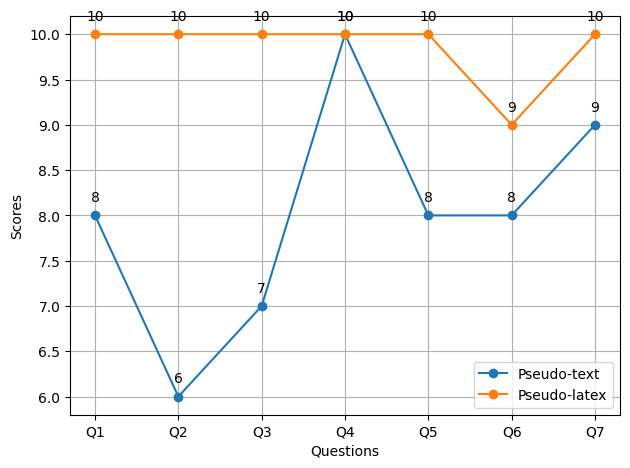}
\caption{DocQA Task Performance Scores Using Pseudo-Text and Pseudo-LaTeX Labels.}
\label{Fig4}
\end{figure}

The document question-answering capabilities of GPT-4 were assessed using pseudo-texts, with performance scores presented in Figure ~\ref{Fig4}. Utilizing pseudo-information allowed the question- answering task to achieve a satisfactory level of performance. However, enhancing OCR models to understand specific symbols and preserve structural integrity remains challenging. Notably, the pseudo-text outperformed expectations in understanding structured data, which resulted from including escape characters and mathematical symbols within the text such as '\textbackslash t', '\textbackslash n', or '\%'. While the pseudo-latex-based approach closely aligns with the ground truth, it faces challenges because the accuracy relies highly on OCR regions and image quality.

\subsection{End-to-end vision-language approach}
\begin{figure}
\includegraphics[width=0.9\textwidth]{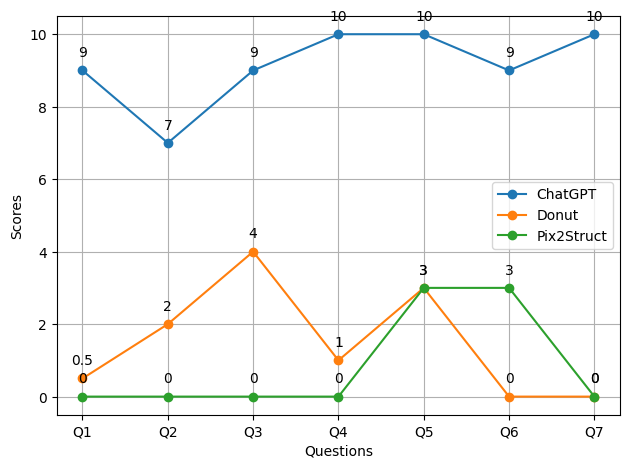}
\caption{Performance Scores of Three pre-trained Models on Enhanced-Image Quality Test Cases.}
\label{Fig5}
\end{figure}

As described in Section 3.3, we evaluate three models: 1) GPT v4 \cite{chatgpt2024}, 2) Donut \cite{kim2022ocr}, and 3) Pix2struct \cite{lee2023pix2struct}. Figure ~\ref{Fig5} illustrates the result scores of pre-trained three models from the enhanced-image quality sample test cases.

\begin{figure}
\includegraphics[width=0.9\textwidth]{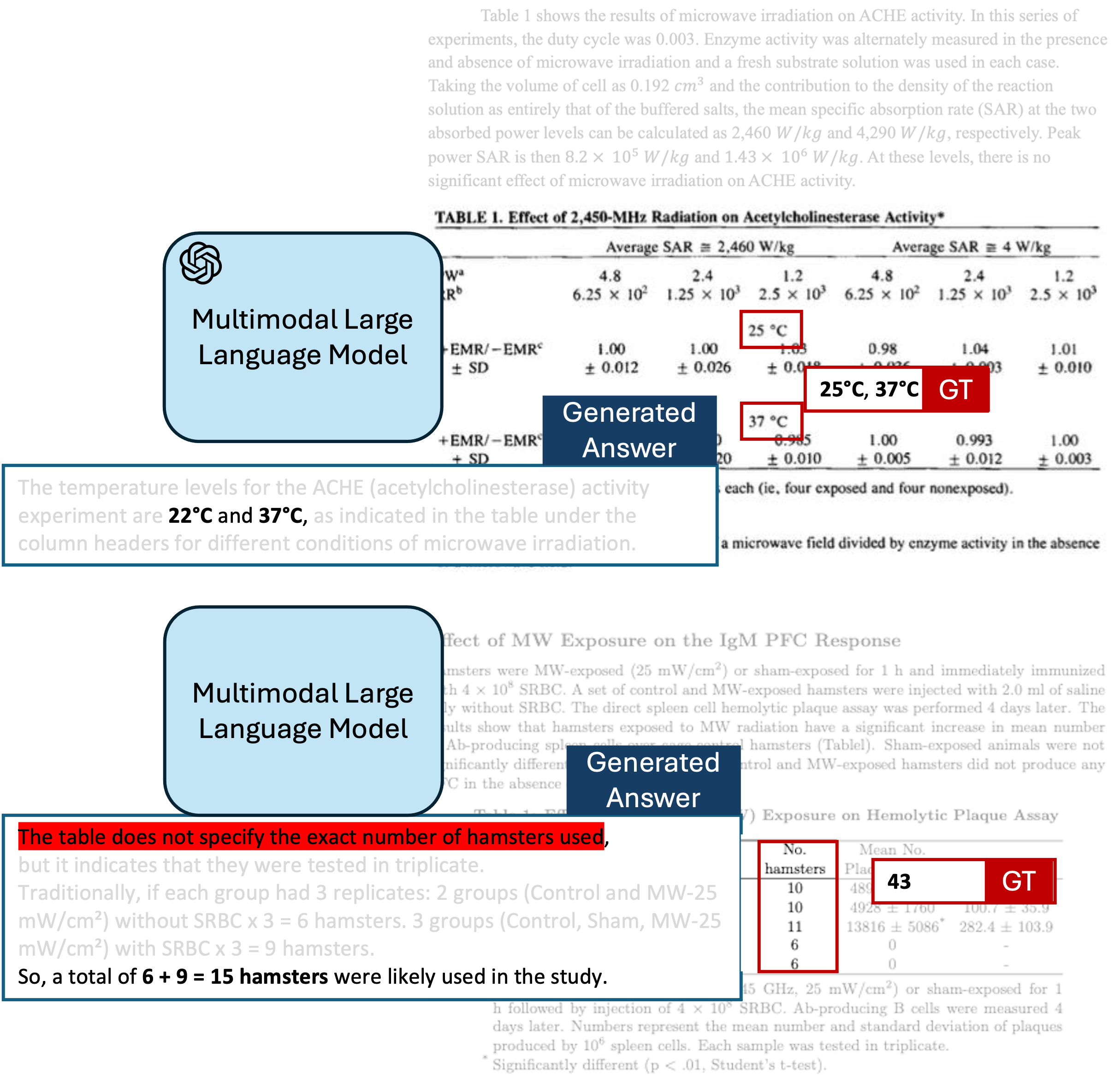}
\caption{Examples of Challenging Cases in GPT-4's Response Generation.}
\label{Fig6}
\end{figure}

GPT-4 demonstrates the highest scores in our sample test cases due to its capabilities as a multimodal large language model trained on large-scale data. However, slight challenges remain where the model generated incorrect answers in Case 1-b. This might be because the model's performance is still susceptible to degradation due to the quality of input images. Figure~\ref{Fig6} highlights error cases in the GPT model and illustrates the impact of poor image quality on performance. In particular, Question 2 often falls under Case 1-b, where the answer is extracted solely from tables. A positive aspect is that ChatGPT appears to understand semantics of mathematical symbols and text, such as temperature symbols, frequency units, and statistical terms. A clear example of this characteristic is demonstrated when it returns 'sample size' in response to the question Q2: 'How many animals were used?', even though it was labeled as 'sample size' in the table.

The Donut and Pix2Struct models perform poorly in our test cases because these models require fine-tuning to understand the specific terms and generate new fixed answers about our data. The reason why Donut slightly outperforms Pix2Struct is that Donut can extract parts of tabular data, whereas Pix2Struct may fail in parsing screenshots of our data, which consists of both text and structured formats.

\subsection{Table structure-aware approach}
\begin{figure}
\includegraphics[width=0.9\textwidth]{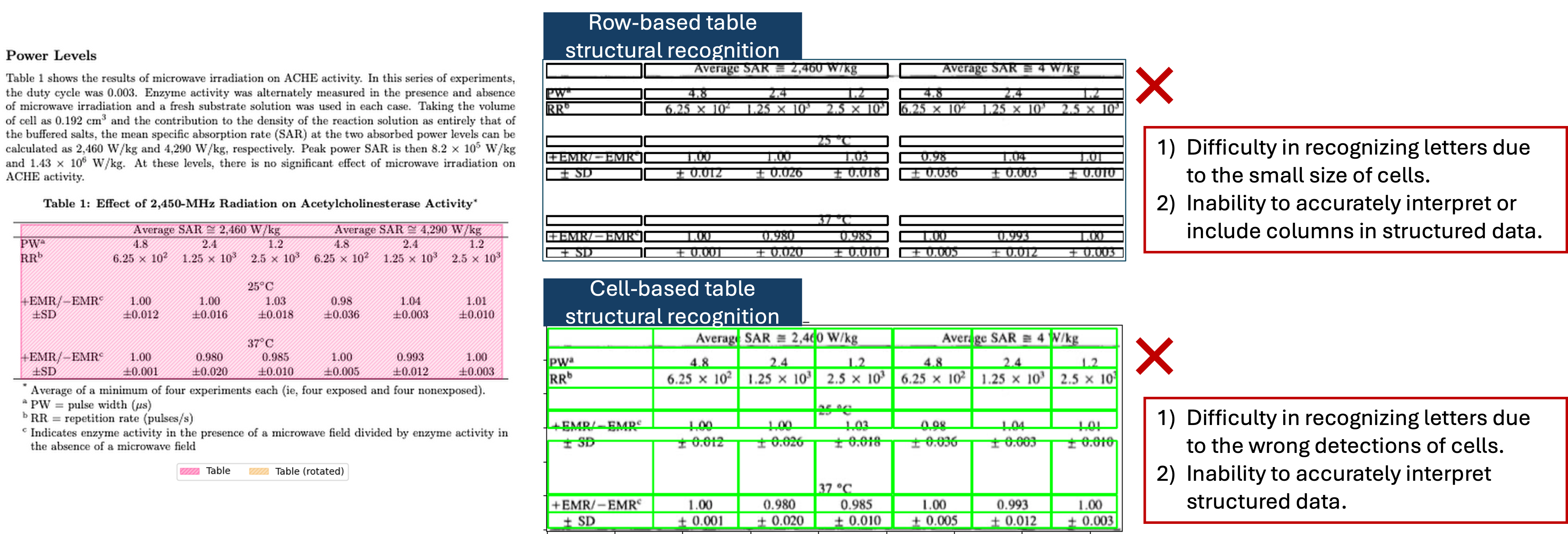}
\caption{The Challenges in the Table Structure-Aware Approach.}
\label{Fig7}
\end{figure}

As described in Section $3.4$, we implement multi-stages of document question-answering by extracting tabular structure information.
The model exhibits two challenges: (1) non-trivial tabular layout and (2) spatially dense structure, as shown in Figure ~\ref{Fig7}.

\subsubsection{Non-trivial tabular layout}
We consider a table to have a trivial layout structure if it features the same number of rows in every column and the same number of columns in every row; essentially, it is a regular grid structure. Conversely, when any row or column deviates by having a different number of columns or rows, respectively, compared to others, we classify it as having a non-trivial layout structure, which is the case here. For instance, if a table includes a merged cell, it is considered to possess a non-trivial layout structure. This poses a challenge to process as merged or irregular cells disrupt standard data extraction and analysis techniques, often requiring specialized algorithms to interpret and organize the information accurately.

\subsubsection{Spatially dense structure}
Aside from the complexity of the layout, the density of the content also adds additional challenges to effective table extraction. This is because dense amounts of text typically cause OCR engines to struggle with text recognition due to closely located non-textual components, such as cell borders.

Given these challenges, the table structure-aware approach relies heavily on the capability of each stage to accurately process and extract information from complex tabular data. However, implementing a multi-stage process is proven to be impractical in handling the intricacies of such dense and non-trivial table structures. The approach faces limitations in maintaining both the accuracy and efficiency required for effective document question answering, suggesting that further refinement and specialized techniques would be necessary to improve performance in these scenarios.

\section{CONCLUSION AND FUTURE WORK}
This study explored three distinct approaches for processing table data in scientific papers to improve the performance of extractive question answering in the context of systematic reviews. Considering the sample testing these methodologies on ten RF-EMF-related documents, our findings reveal that they face significant challenges in preserving structural and contextual information, especially in complex and degraded documents, while OCR-based methods are foundational. Pre-trained model models such as Donut and Pix2Struct demonstrate strong potential, particularly in handling document images directly without relying on traditional OCR. However, these models require fine-tuning to address the diverse and intricate nature of scientific tables.
The table structure-aware approach, though essential for maintaining the integrity of tabular data, encounters difficulties with non-trivial layouts and densely packed information, underscoring the need for further refinement. Our future research is aimed at fine-tuning these models using augmented datasets that can better represent the diversity of document layouts and symbols encountered in scientific literature. Moreover, integrating OCR location data and refining the processing stages in table recognition could significantly enhance the accuracy and reliability of extractive question-answering tasks, thereby advancing capabilities of systematic review tools.

\section*{Ackowledgements}
This work was partially supported by an Institute of Information communications Technology Planning Evaluation (IITP) grant funded by the Korean government (MSIT) ($2019-0-00102$, A Study on Public Health and Safety in a Complex EMF Environment). This work was also supported by Institute of Information \& communications Technology Planning \& Evaluation (IITP) grant funded by the Korea government (MSIT) ($No. RS-2024-00466966$, A systematic study on health risk of EMF exposure in advanced wireless service environment.   


\bibliographystyle{ACM-Reference-Format}
\bibliography{references} 

\end{document}